\documentclass[authoryear,12pt,1p]{elsarticle}
\usepackage{graphicx}

\journal{New Astronomy Reviews}
\begin{document}

\begin{frontmatter}


\title{3D spectroscopic study of galactic rings: formation and kinematics}

\author[avm]{A.V. ~Moiseev\corref{cor1}}
\ead{moisav@gmail.com}

\author[dvb,dvb2]{D.V. ~Bizyaev}
\ead{dmbiz@apo.nmsu.edu}

\cortext[cor1]{Corresponding author}
\address[avm]{Special Astrophysical Observatory, Nizhnij Arkhyz, 369167 Russia}
\address[dvb]{Apache Point Observatory, Sunspot, NM, USA}
\address[dvb2]{Sternberg Astronomical Institute, Moscow, 119992, Russia}

\begin{abstract}
In this review we consider various ring structures that are observed in galaxies.
Formation and evolution of the rings are interesting problems in 
studies of galactic morphology. They are related to such fundamental aspects of galactic evolution and dynamics as the nature and distribution of the dark matter in galaxies, galactic interactions
and internal secular evolution of galactic substructures. A significant fraction of galactic rings forms in the disks due to gravitational torques from bar-like patterns. In contrast to this internally driven origin, the phenomenon of the polar-ring galaxies is closely connected with the
processes of intergalactic interactions and merging. A rare class of collisional rings reveals the
density waves triggered in the stellar and gaseous disks after a strong
head-on collision with a companion. We briefly review the status of studies of gas kinematics
in the rings of different origin. We stress that velocity fields of the ionized gas obtained with the
Fabry-P\'{e}rot interferometers provide a very important information for these
studies.
\end{abstract}


\end{frontmatter}

\section{Introduction}

Direct imaging reveals high- and low-contrast annular structures in numerous
galaxies. Rings are mostly observed in barred galaxies but also can be
found in some unbarred and early-type galaxies. The linear size of these structures
ranges from dozen pc for the ultracompact nuclear rings \citep{comeron08} up to
100 kpc in giant colliding systems \citep{ugc7069}.  The rings are
often found to be the sites of active star formation and, in majority
of cases, results
of some collective processes in stellar-gaseous disks associated with
resonance phenomena or annular density waves. \citet{butacombes96} presented
an extensive review of ring phenomenon, first of all concerning such
structures associated with bars or other non-axisymmetric disk features.
In general case, rather different mechanisms are responsible for
the formation of the inner and outer galactic rings. The galactic rings can be classified as follows:

\begin{enumerate}
\item A large class of the rings which is connected with the Lindblad's
dynamical resonances in the galactic disks (`resonance rings').
\item The objects  formed  after a strong head-on collision with a companion
(`collisional rings'). The prototype is the Cartwheel galaxy.
\item Polar rings are also results of the galaxy-to-galaxy  interactions or
accretion of matter with a roughly orthogonal direction of its angular
moment with respect to the target galaxy. \citet{butacombes96} mentioned a
type of `accretion rings' which includes
also objects with a coplanar orientation of the ring and main galaxy.
The Hoag's object is a possible prototype.
\end{enumerate}

\begin{figure*}
\centerline{
\includegraphics[width=5.5 cm]{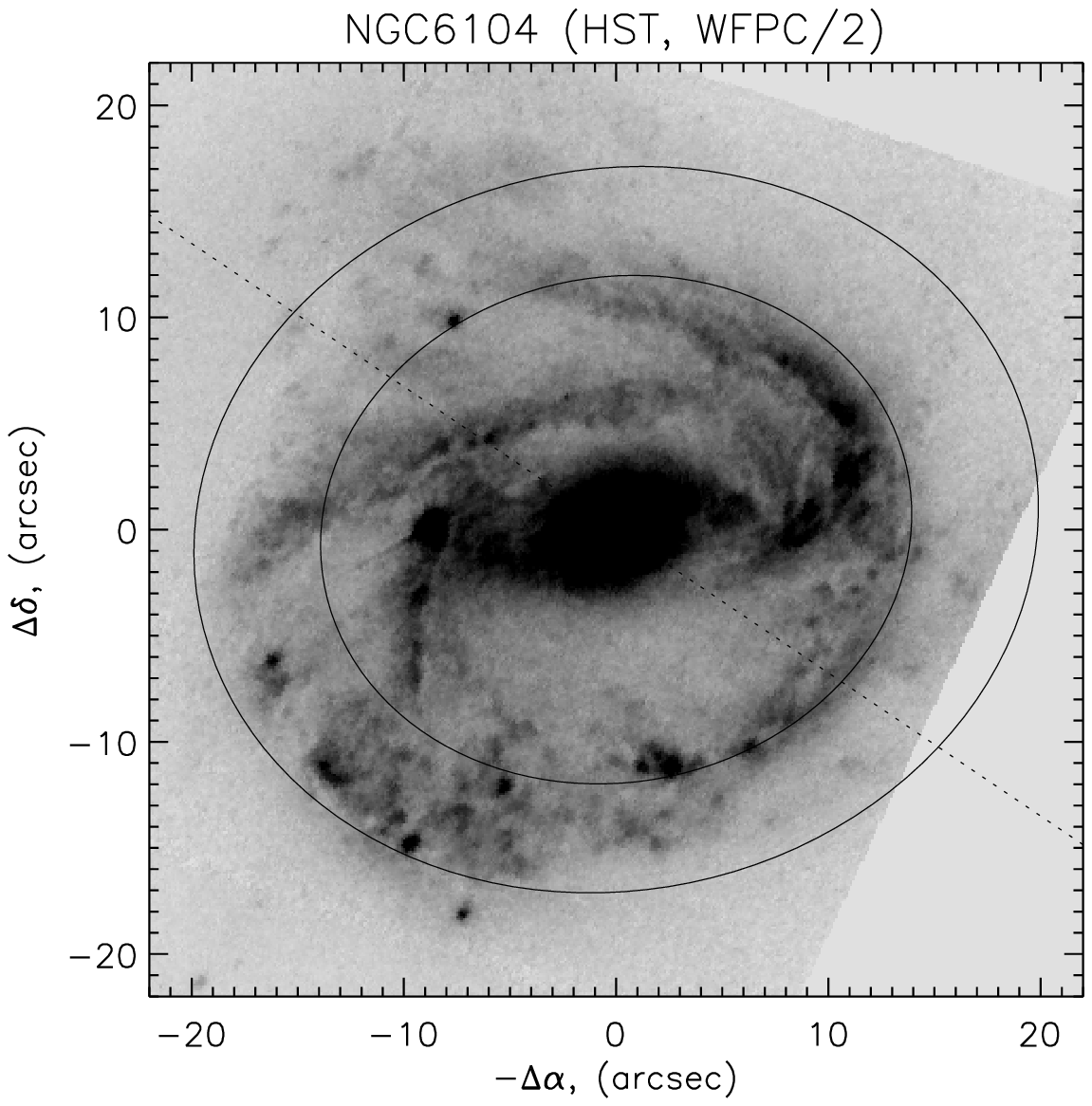}
\includegraphics[width=5.5 cm]{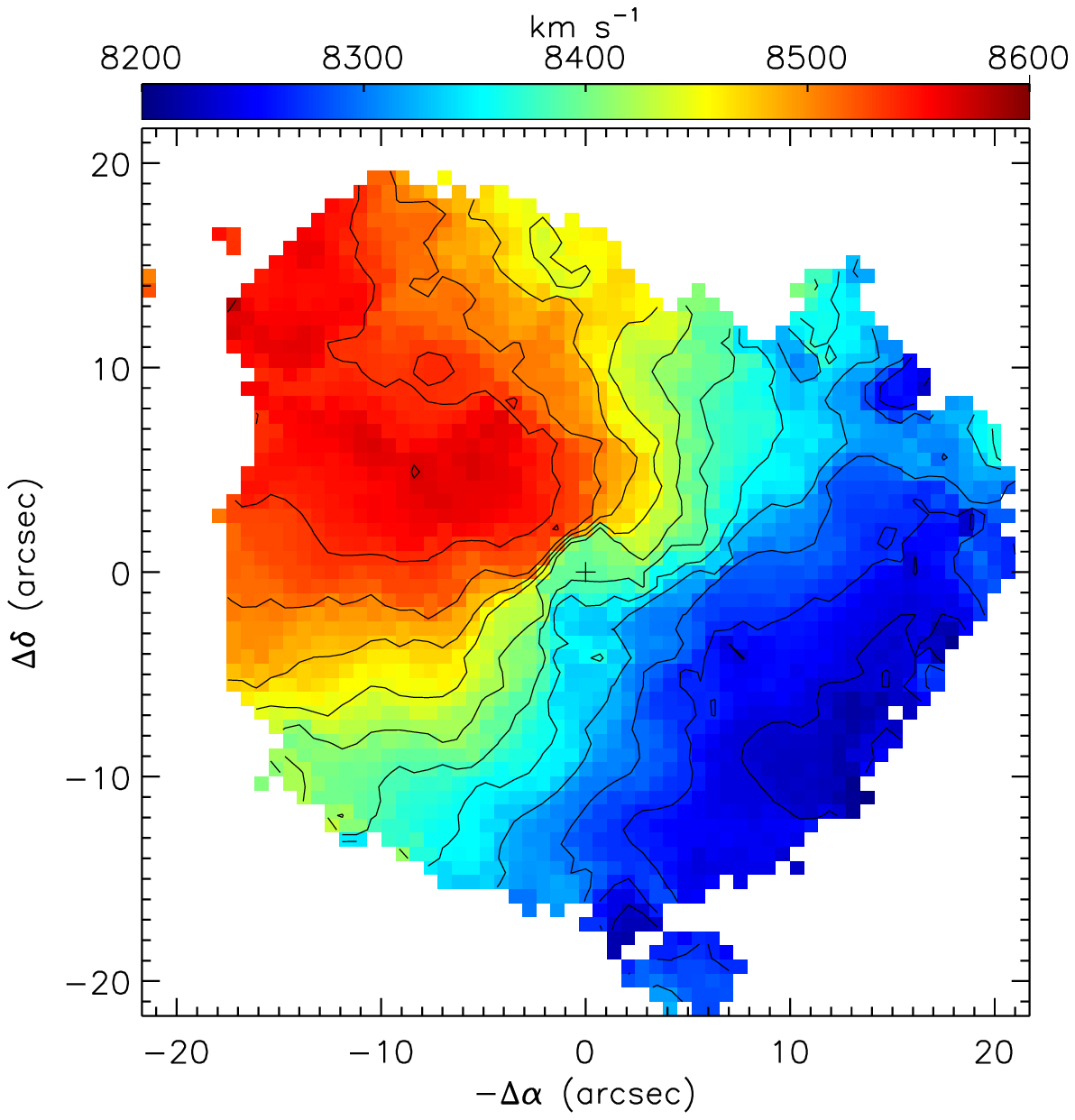}
\includegraphics[width=5.5 cm]{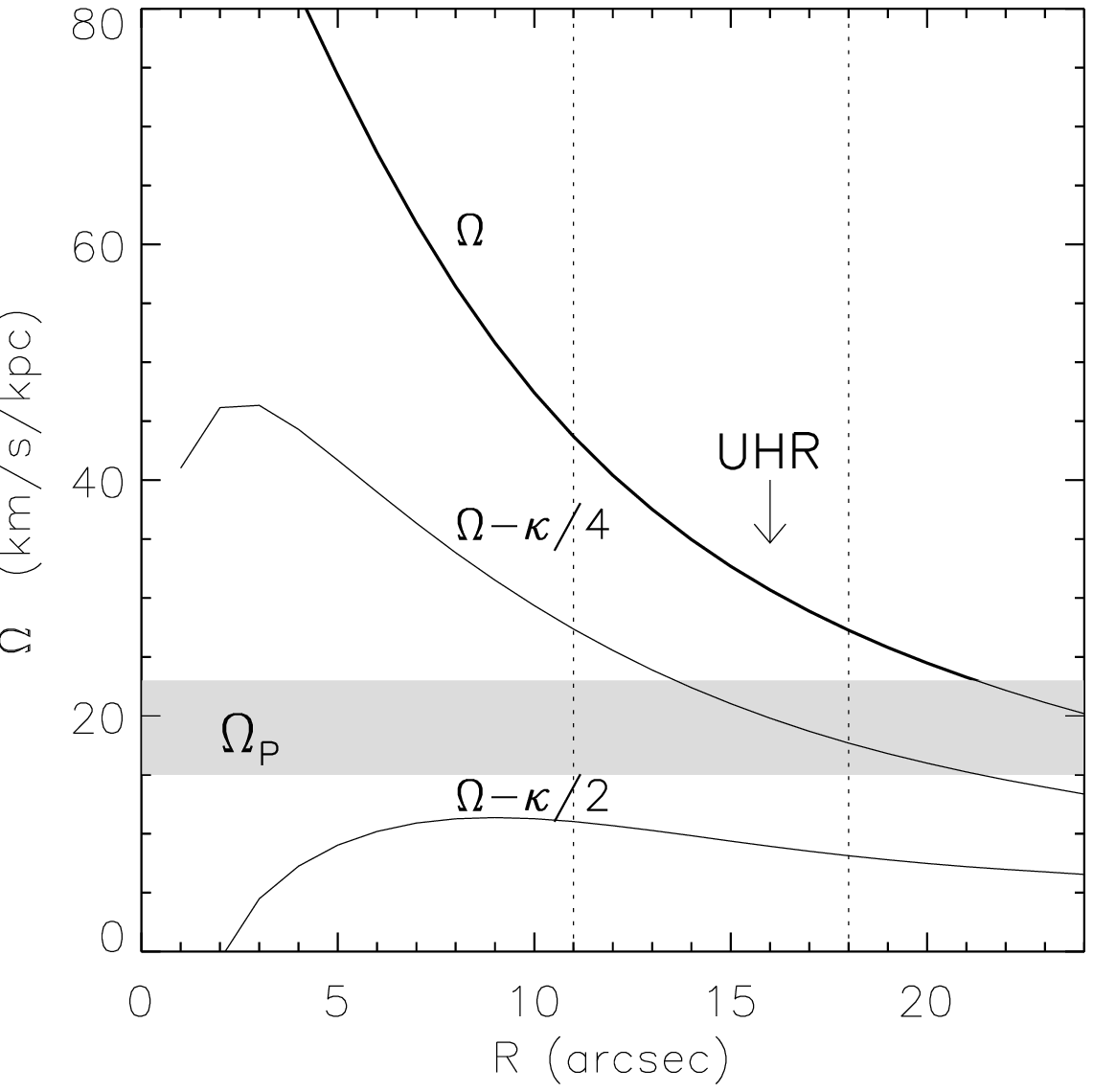}
}
 \caption{
NGC~6104. Left: The HST/WFPC2 image deprojected to the face-on position. The
dotted line corresponds to the line of nodes, the ellipses designate the
ring's borders.
Middle: the   velocity field  in the  H$\alpha$ emission line derived from
the SCORPIO/FPI data.
Right: the diagram of characteristic frequencies. The gray
color designates the range of the pattern speed of the bar ($\Omega_p$), the
dotted lines
indicate the maximum and minimum ring radii.}
\label{fig1}
\end{figure*}

The origin and evolution of the rings are closely connected with various
aspects of galactic life. A comprehensive study of evolution of the
rings should involve various observational data, among which the
kinematical information is one of the most important. A panoramic (3D)
spectroscopy which provides spectra for each spatial element of certain
two-dimensional field of view is a powerful technique in the spatial-resolved
study of kinematics in the galaxies. In this article we briefly review
some results of studies of inner kinematics in the ring galaxies conducted
mostly with the
scanning Fabry-P\'{e}rot interferometers (FPIs).  The FPI data allow to
obtain the
large-scale velocity fields, maps of the velocity dispersion and
monochromatic flux
in the ionized gas simultaneously. The main goal of such observations is
reconstruction, under various model assumptions, of spatial distribution
of the gas velocities.  Usually FPI yields the angular resolution significantly
better than that obtained with the radio observations of neutral and
molecular gas. This is critical when the kinematically decoupled regions are considered.
Majority of the results presented in this review were obtained during last decade with
Russian 6-m telescope and a multi-mode spectrograph SCORPIO \citep{Afanasiev2005}.
Also we mention some results of studies of the ring kinematics obtained
by several other teams that use FPI and/or integral-field devices.

\begin{figure*}
\centerline{
\includegraphics[width=5.5 cm]{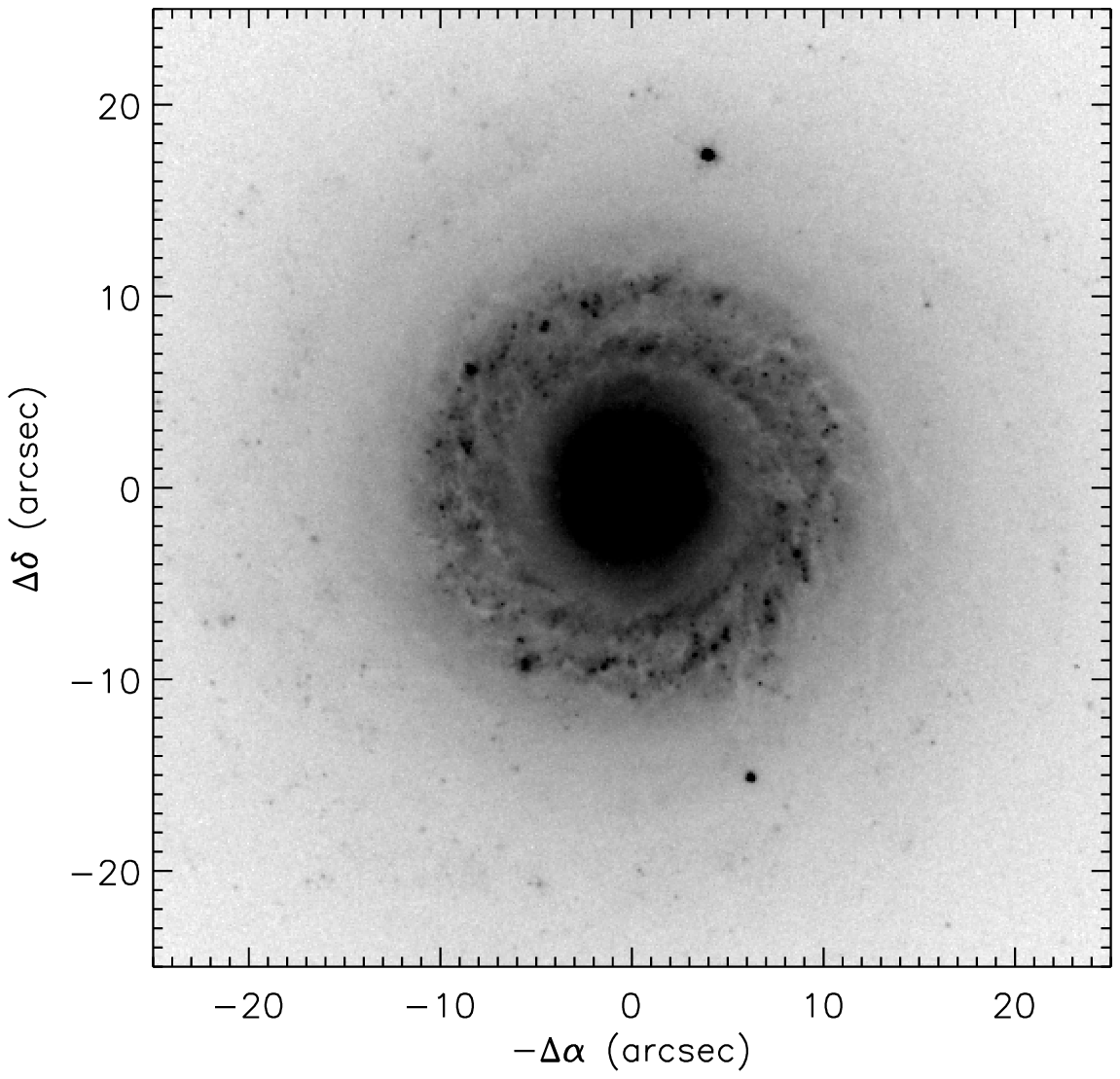}
\includegraphics[width=5.5 cm]{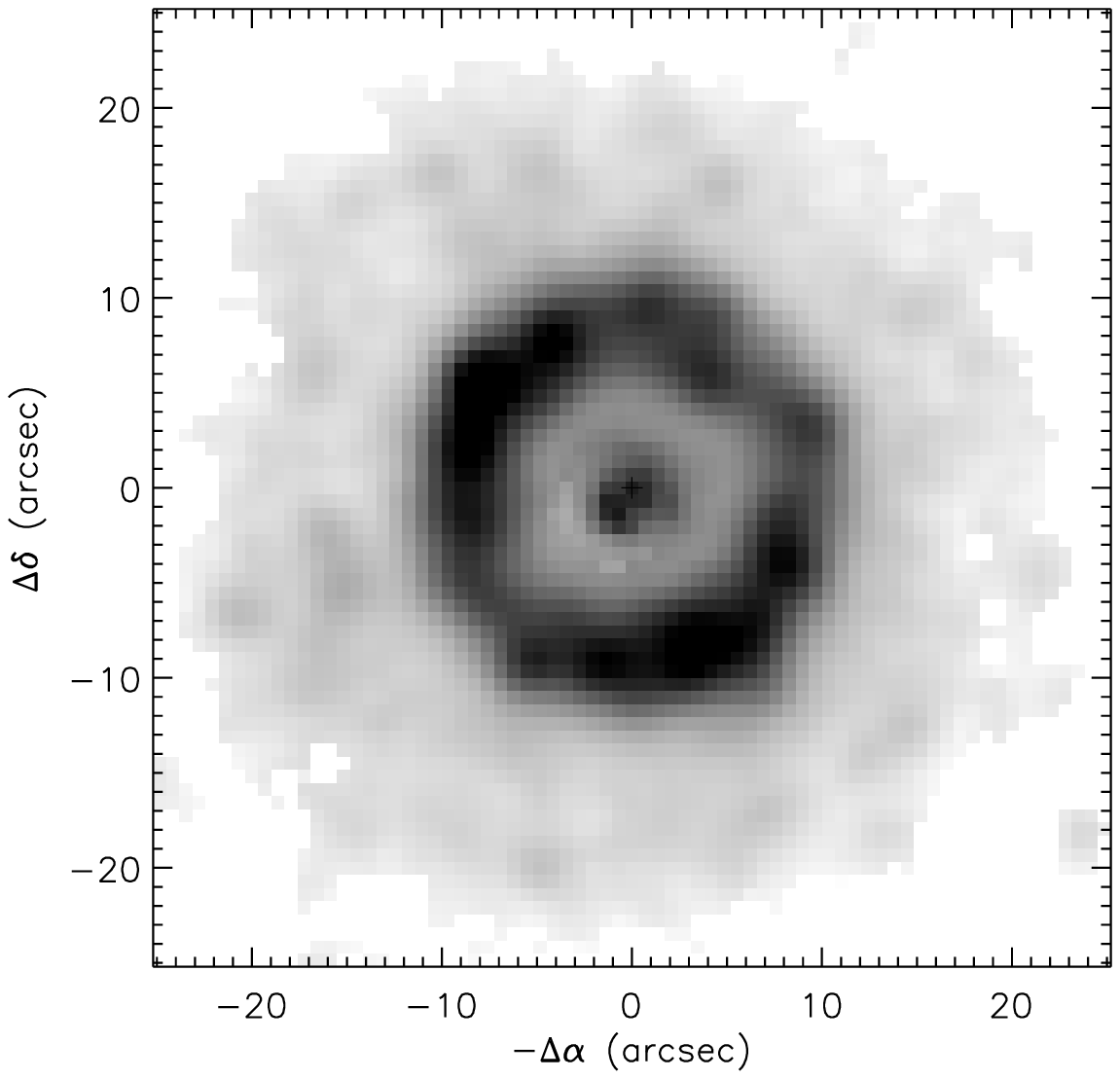}
\includegraphics[width=5.5 cm]{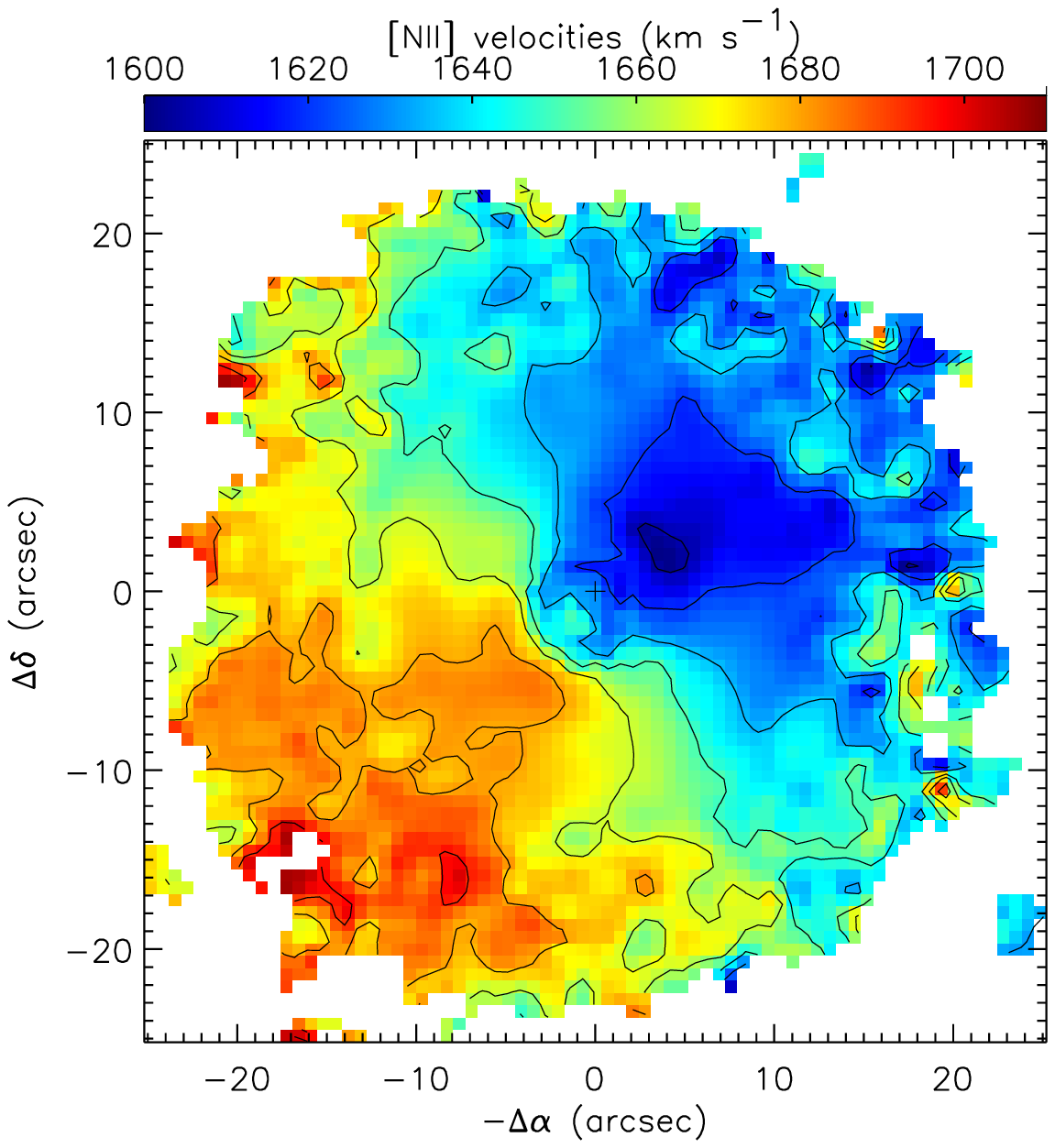}
}
 \caption{
NGC~7742. Left: a HST/WFPC2 image. Middle: a monochromatic image in the [NII]
emission line.  Right: the ionized gas velocity field \citep{silmois06}}
\label{fig2}
\end{figure*}

\section{Resonance rings}
\subsection{A classical picture}

According to the different statistical estimations, the
rings are observed in a
significant fraction, up to 20--30\%, of spiral and lenticular galaxies and
closely associated with the non-axisymmetric structures like bars, ovals or
triaxial bulges. This fact is traditionally explained with the
orbits crowding and
the gas accumulation at the Lindblad's resonances
(see a review by \citet{butacombes96}). In the terms of the theory, the empirical
subdivision of the rings to the `nuclear', `outer' and `inner'
corresponds
to the inner (ILR), outer (OLR) and ultraharmonic (UHR) dynamical
resonances between the epicyclic oscillations in the stellar component
and rotation of the bar.
Kinematical effects observed in the resonance rings are not very prominent
because the amplitude of possible non-circular gas motions (tens of
$\mbox{km}\,\mbox{s}^{-1}$) is significantly smaller than velocities of the
circular rotation
($150-200\,\mbox{km}\,\mbox{s}^{-1}$).
It is very difficult to extract the pure rings kinematics
itself. In general, it is a part of complex inward/outward motions driven by
the bar potential. Interpretation of the line-of-sight velocities
is often difficult because their projection depends on the
assumed orientation of the galaxy disk. The high-accuracy FPI velocity field
in
combination with a good coverage of the filed of view (thousands of
independent measurements per typical field) helps to understand the overall
picture of peculiar
gas motions. Long history of study of the bright star-forming ring
in NGC~4736 is a
good illustration. Firstly, \citet{vanderKruit1974} from long-slit spectra
suggested the idea of `expanding H~II regions ring'.  Than, using the FPI data
\citet{Buta1988} concluded that the ring kinematics can be better interpreted as
a result of the secular evolution in this galaxy due to its ILR. Using new
TAURUS-II FPI observations \citet{Casiana2004} drawn a more detailed picture
which includes a radial stream driven by the bar and chaotic gas motions
associated with
starburst knots in the galaxy. \citet{Moiseev2004} also detected the inflow
gas motions related to the bar in this galaxy.

The complex morphology of the rings$+$bar$+$spiral structure is convenient to
study in moderately inclined galactic disks.
Restoration of the rotation velocity fields from observations in the case of
nearly face-on
orientation is a serious problem which can be resolved via examination
of the velocity fields under assumption of domination of the rotational
component. A general analysis of both the rotation curves extracted from the FPI
velocity fields and from the surface brightness distribution for
several galaxies (each of them has more than one ring) shows that the
loci of the rings agree very
well with predictions from the dynamics of barred galaxies
\citep{Arsenault1988,Buta1995,Buta1998,ButaPurcell1998,Fathi2007}. This
way allows to estimate the angular speed of the bar pattern
($\Omega_p$) which is one of the key characteristics for understanding
the galaxy's secular evolution. The ionized gas velocity fields and
observed morphological features were reproduced by numerical simulations in
the works which considered IC~4214 \citep{Salo1999} and NGC~6782 \citep{Lin2008}.
The temporal evolution of their disks was studied and the pattern
speed was restored as one of the main parameters in the simulations.

A model independent method of $\Omega_p$ estimation
introduced by \citet{TW1984} came into a wide usage
in studies of the stellar component in early type galaxies
and of the molecular disks in late type galaxies.
However, some authors \citep{Hernandez2005,Beckman2008} show that the T-W
technique (with some modification) works correctly also for the
ionized gas which is an example of
a media where the continuity equation gets broken. A barred Seyfert
galaxy NGC~6104 is a good example for it \citep{Smirnova2006}.
The pattern speed derived from the H$\alpha$ velocity field agrees very well
with the ring location near 1:4 ultraharmonic resonance, as it follows from the
theory (Fig.~\ref{fig1}). The $\Omega_p = const$ line in Fig.~\ref{fig1}
does not intersect the
$\Omega- \kappa/2$ curve, therefore the bar has no ILR where the radial gas
motions toward the center cease, so gaseous clouds can reach the center
and provide a fueling for the nuclear activity.

\subsection{Rings without a bar}

The objects which are failed to be explained in the frames of a simple theory of
dynamical resonances are very interesting. There are several examples of
spectacular nuclear rings in unbarred galaxies.  Most of these galaxies are
observed face-on, so the conclusion about the bar absence is rather solid
for them.
Explanations for the origin of such rings include the resonance effects
produced by a weakly triaxial potential employ an idea that such rings
could be
explained by weakly triaxial distortions, or by tidal actions from
a gravitationally bounded
companion, since the non-axisymmetric gravitational perturbations in these
cases are similar to those from a
bar (see \citet{butacombes96} and \citet{silmois06} for references).
\citet{Knapen2004} suggest that the latter mechanism, i.e. a recently
occurred interaction with
a dwarf gas-rich satellite, is able to produce the nuclear ring in the
unbarred galaxy
NGC~278. Their conclusion is based on the strongly peculiar kinematics of
the neutral and ionized hydrogen detected in the observed gas velocity fields.
\citet{silmois06} proposed the same scenario for the emergence of the
rings in NGC~7217 and
NGC~7742 (fig.~\ref{fig2}). Both galaxies demonstrate some signatures of
recent
minor-merging events in their stricture and kinematics, like the
two-tiered stellar
disks and mutually counterrotating subsystems. NGC~7742 has all its gas in
the counterrotation with respect to the stellar component.

Using results from the integral-field spectroscopy, \citet{Mazzuca2006} also
show an evidence for the tidally induced origin of the ring in NGC~7742.
This idea agrees
with results of numerical simulations of the density re-distribution
during the galactic tidal interactions in the case if a satellite moves
in the plane of the main galaxy in the direction reverse to the main rotation
\citep{Tutukov2006}.

\section{Collisional rings}

Collisional ring galaxies are results of nearly central, head-on passage of a
satellite through the main, or target galaxy. Significant part of all young
stars in such ring galaxies is observed in their star forming rings,
and the galaxies in a total
demonstrate moderately high star formation rates,
like 18 M$_{\odot} \, yr^{-1}$ in the Cartwheel \citep{mayya05},
which is often called a prototype for this class of objects.
They can be subdivided by morphological classes
RE, RN or RK, named after the empty rings, rings with nuclei, or with
dominant knots by \citet{theys76}.

The collisional rings are
rather rare among all ring galaxies. Thus, \citet{Madore09}
in their new {\it Atlas and Catalog of collisional ring galaxies} identified
127 rings of collisional origin from some 7000 peculiar objects.
The rings classification, possible scenarios of their formation and
evolution were considered by \citet{appleton96}
and modeled numerically by \citet{lynds76}, \citet{theys77}, \citet{struck93},
\citet{mazzei95}, \citet{athanassoula97},
\citet{korchagin98}, \citet{vorobyov03}, \citet{bizyaev07}, and by many others.
All authors notice the relatively short lifetime of the collisional rings
(of the order of a few hundred million years) and their transient nature.
Among the three most popular explanations of the collisional
ring supporting mechanisms:
the material wave expanding inside out, or wave of self-inducing star
formation, or propagating density wave (without a real mass transfering),
the latter gets the most reliable support from observations.

The collisional origin of the rings assumes that the ring pattern in the
galaxies is not a steady structure; it should propagate through the disks.
This
gives us an interesting possibility to study the star formation when
its different episodes are separated in time and space simultaneously
since the moving
front of the star formation leaves behind the progressively aging stellar
population (see \citet{bizyaev07}). One of the key ingredient necessary for
understanding the mechanism[s] of
the ring formation and evolution is the picture of the kinematical motions
of the gas and stars in the ring galaxies.

The expansion of the collisional rings was predicted by numerical
simulations, but the ultimate confirmation of the expansion comes from the HI
radio- and H$\alpha$ optical observations focused on kinematics of the
neutral and ionized gas. That makes the 3D  spectroscopy an extremely
important tool in studies of the collisional rings.

The kinematical studies of the gas motion in Arp~10 and the Cartwheel
were performed by \citet{fosbury77},
\citet{charmandaris96}, \citet{higdon96}, \citet{amram98}. They
gave out rather contradictory results on the expansion speed of the
Cartwheel external
ring: 13-30, 53, and 89 km/s from the IFP H$\alpha$, VLA HI and
long-slit spectroscopy, respectively.
Even more discrepancy emerges when the present-day ring expansion
speed is compared with the secular (averaged over last few hundred Myrs) speed
of the ring inferred from the color distribution \citep{vorobyov03},
and the radial distribution of spectral indices \citet{bizyaev07}.
Such a difference finds its explanation in the wave of density model of the
collisional ring expansion. One more argument for this scenario
is a lack of powerful shockwaves at the outer edge of the rings, as it is
noticed for the case of Arp~10 \citep{bizyaev07}.

The   FPI data, in a combination with the long-slit spectroscopy and
photometry, can help to effectively investigate the history of propagating
star formation in the ring galaxies, as it has been done for Arp~10.
The H$\alpha$ velocity field obtained by \citet{bizyaev07} shows an evidence
for significant radial motions in both outer and inner galactic rings.
In \citet{bizyaev07}, we fit the kinematical data with a
model velocity field taking into account the circular rotation, possible
radial motions and projection
effects.  The expansion velocity in the NW part of the outer ring in Arp~10
exceeds
$100\,\mbox{km}\,\mbox{s}^{-1}$, whereas it attains only $30\,\mbox{km}\,\mbox{s}^{-1}$ in the SE part. This asymmetry may
be a result a substantially off-centered passage of the satellite-intruder
through the plane of the main galaxy.
We identified the intruder as a former early-type spiral galaxy
whose mass was about $1/4$ of that of the target galaxy before their
collision. The
size of the inner and outer rings, their expansion velocity
and the radial profile of the gas circular velocity can be reproduced
in our model of a rather off-center collision (the impact occurred at about
3 kpc from the nucleus of the target galaxy) which happened
approximately 85 Myr ago.

\begin{figure*}
\centerline{
\includegraphics[width=5.5 cm]{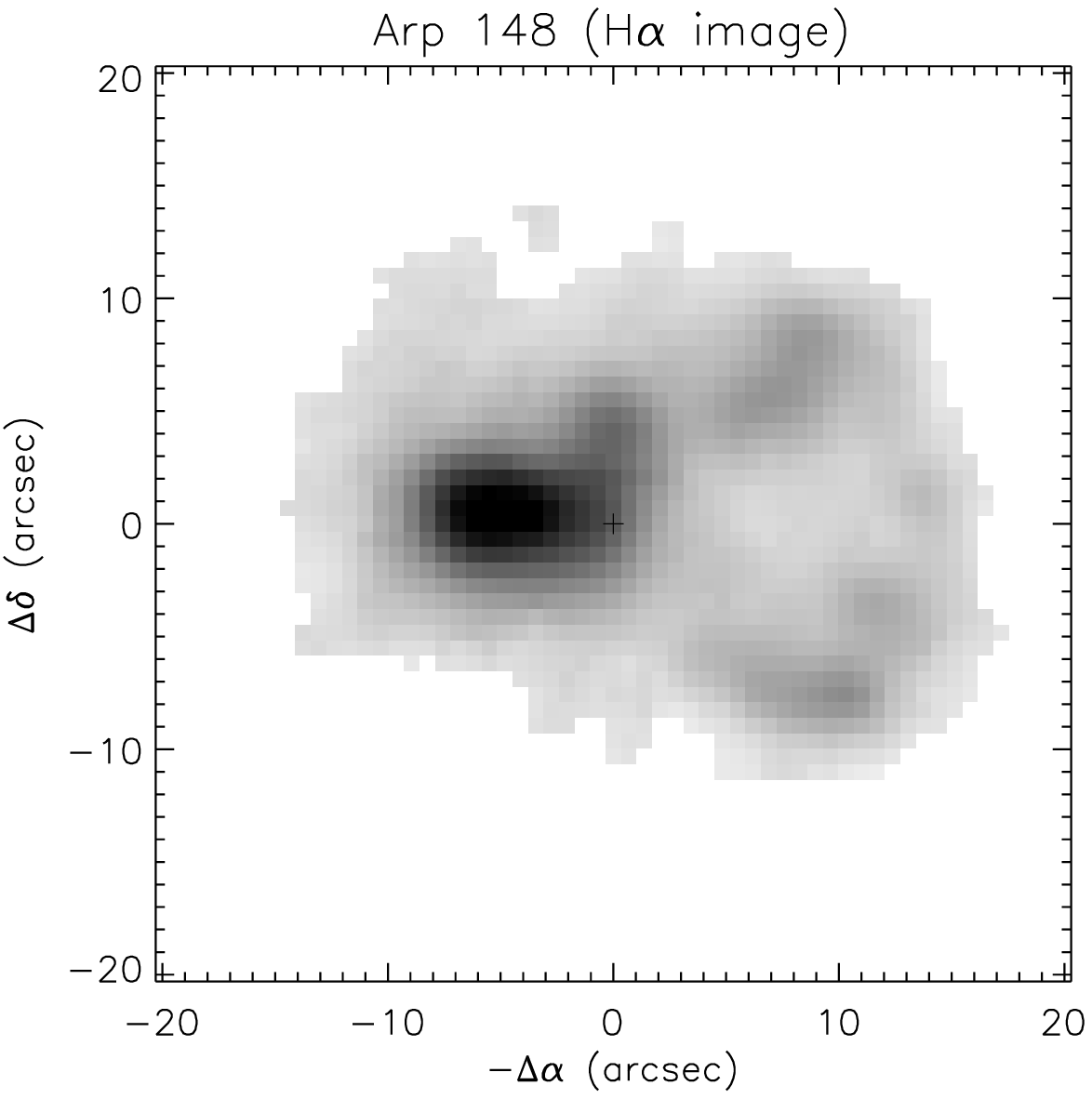}
\includegraphics[width=5.5 cm]{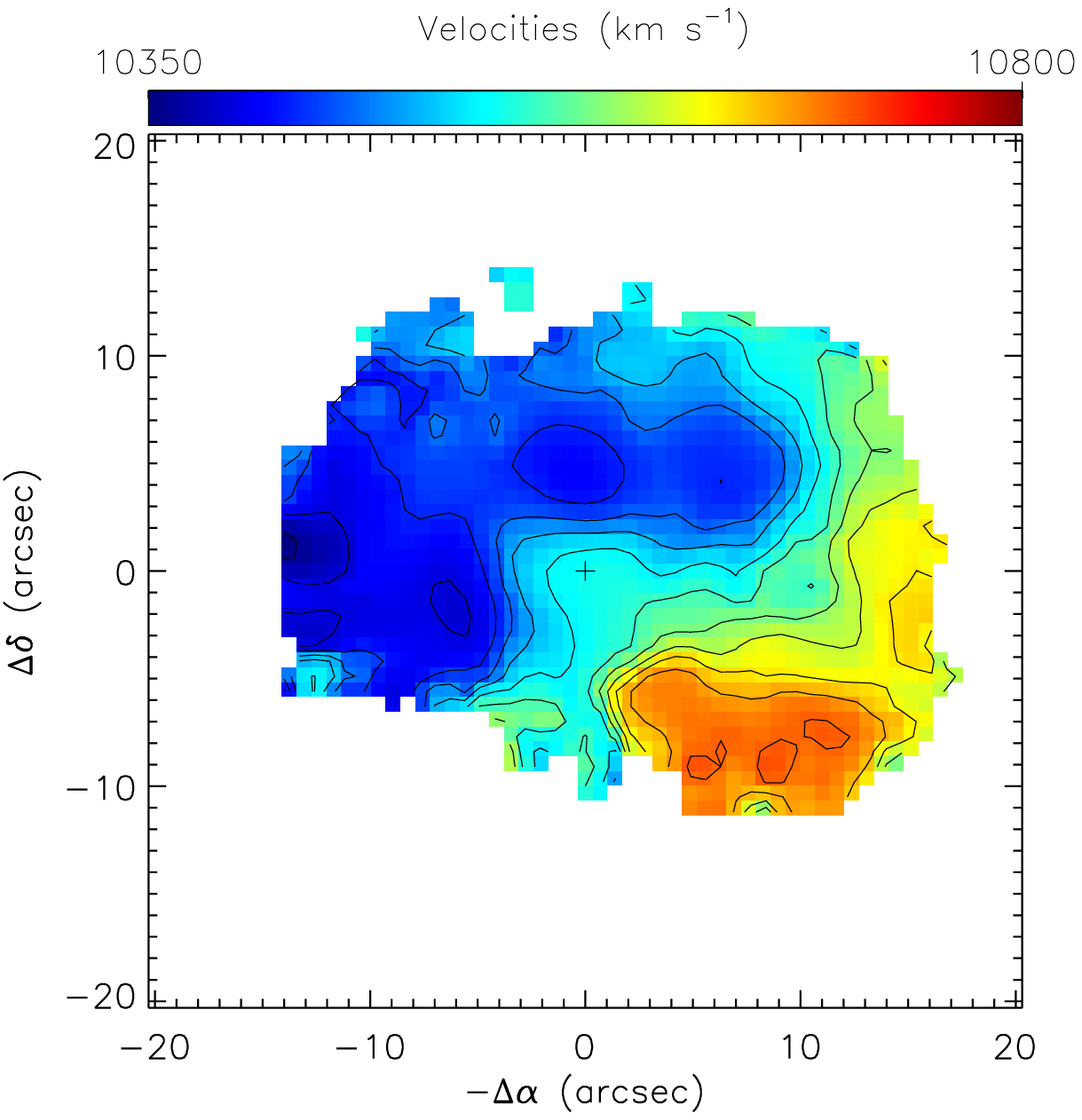}
\includegraphics[width=5.5 cm]{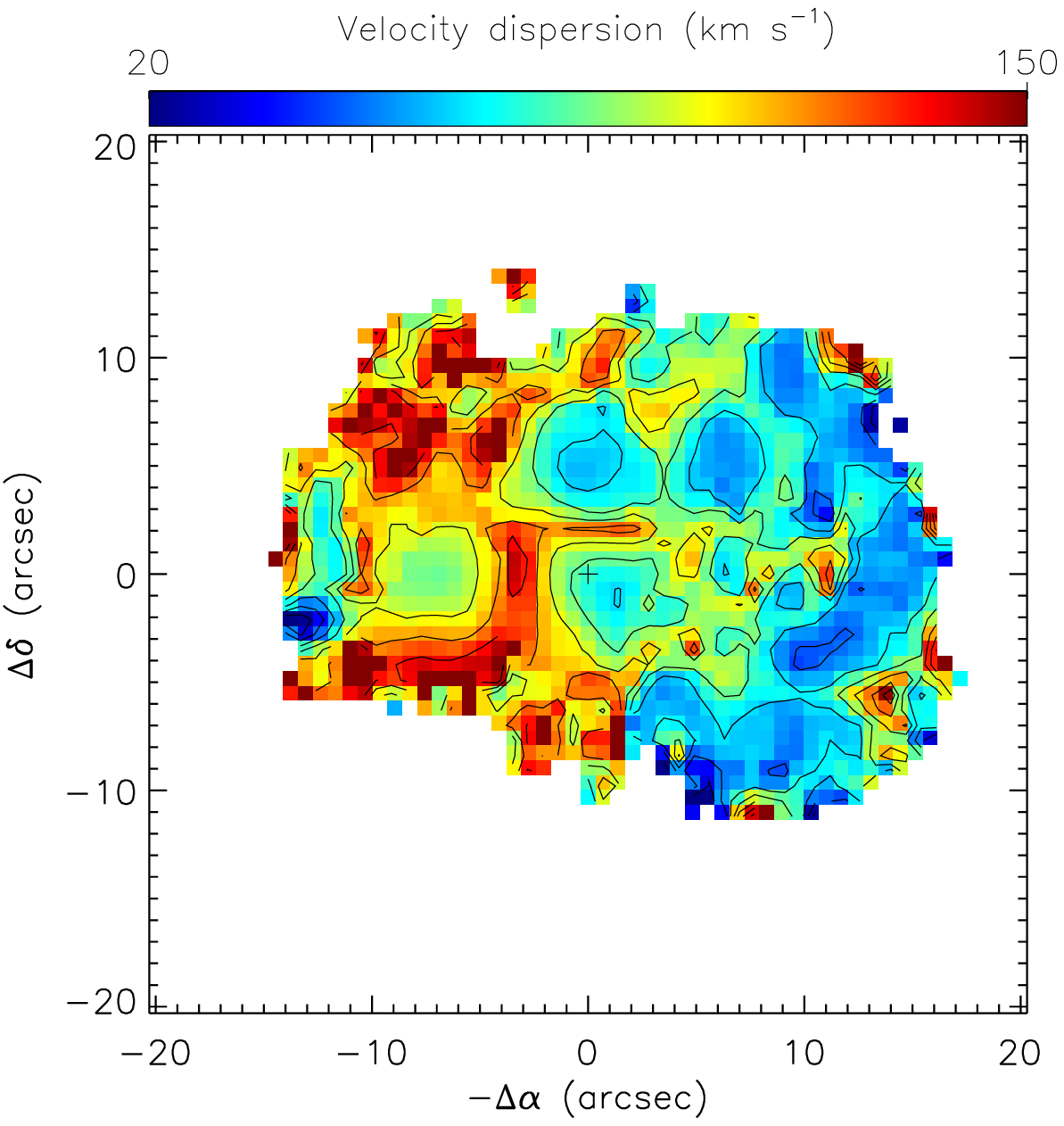}
}
\centerline{
\includegraphics[width=5.5 cm]{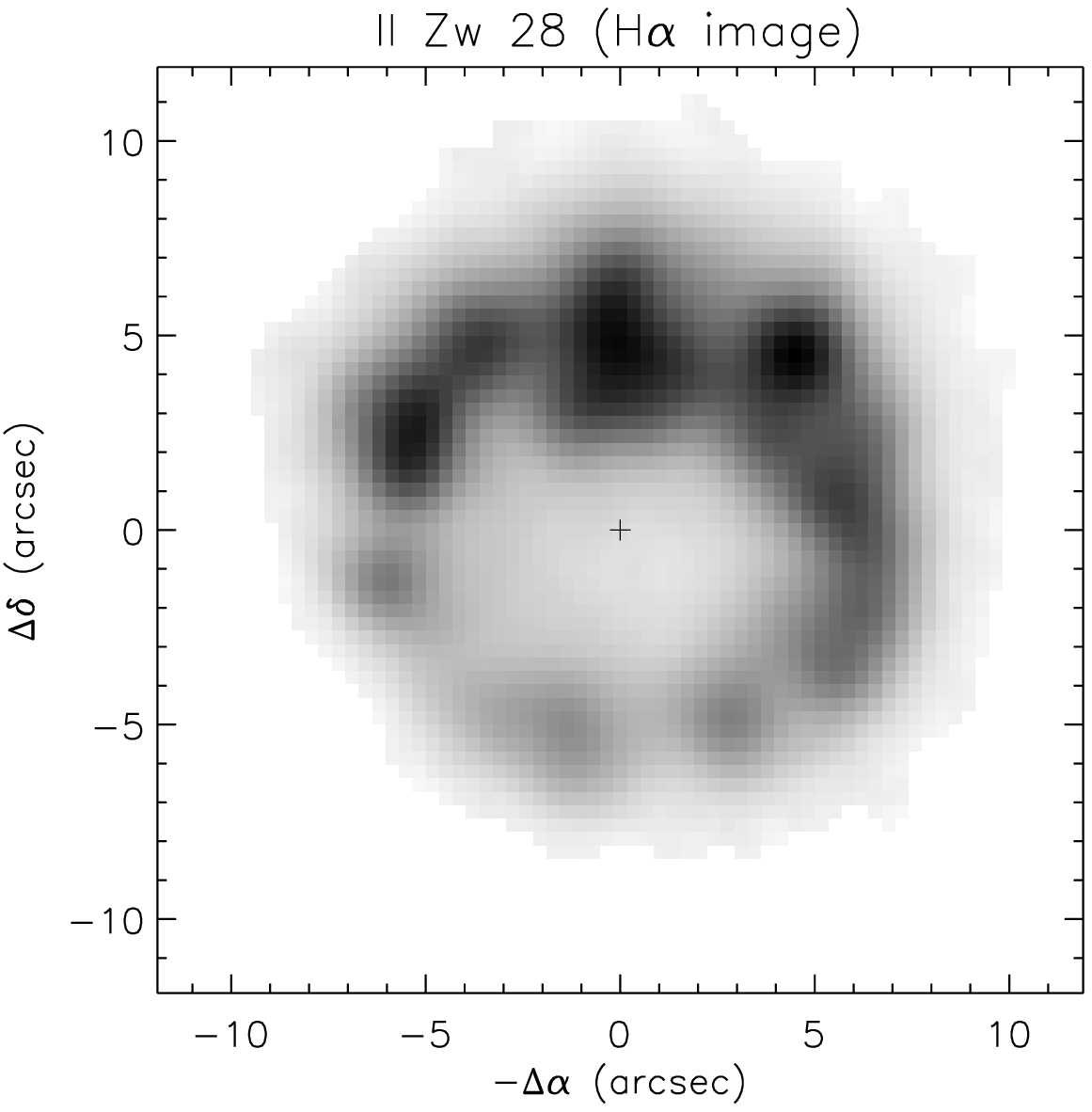}
\includegraphics[width=5.5 cm]{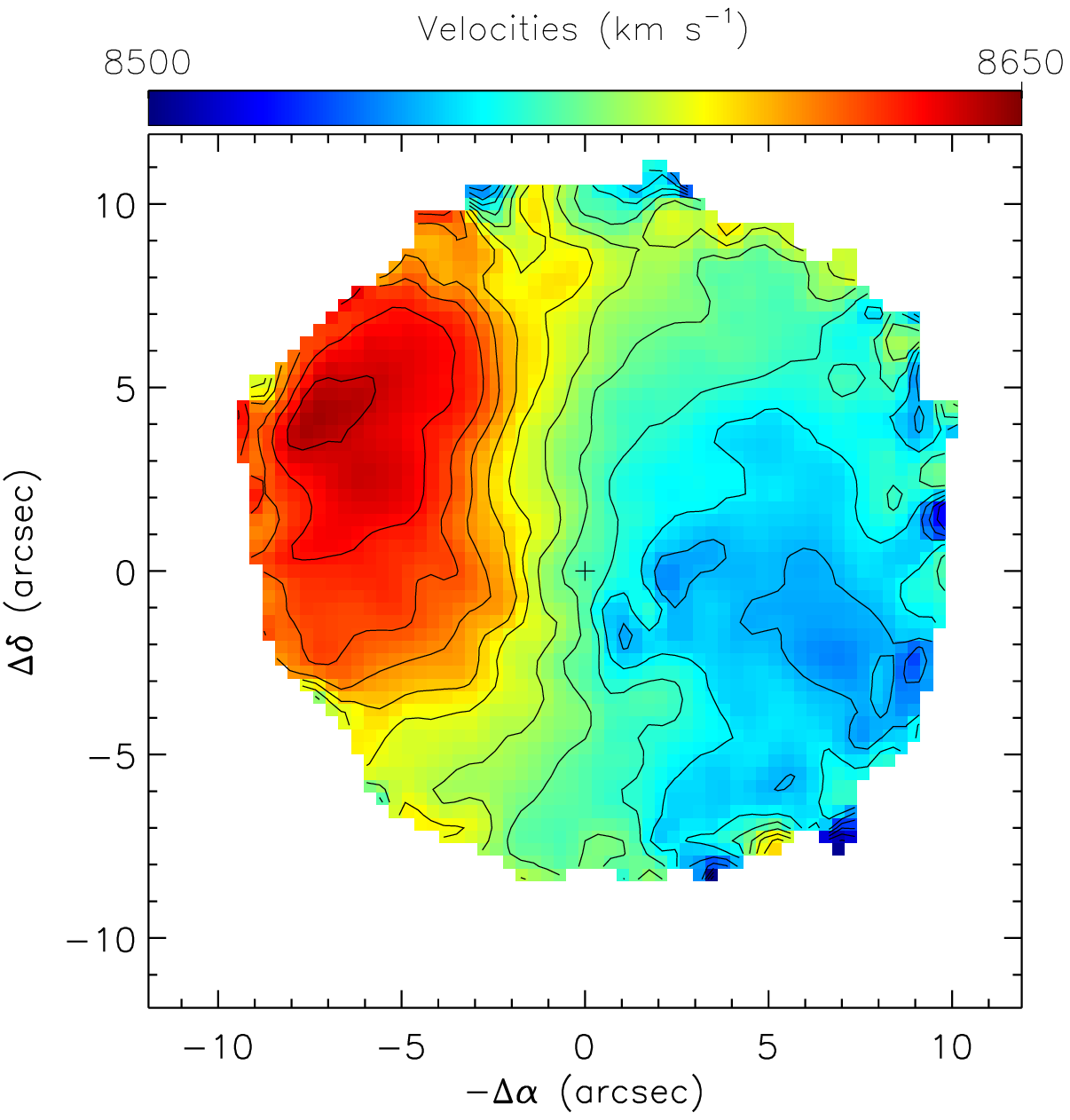}
\includegraphics[width=5.5 cm]{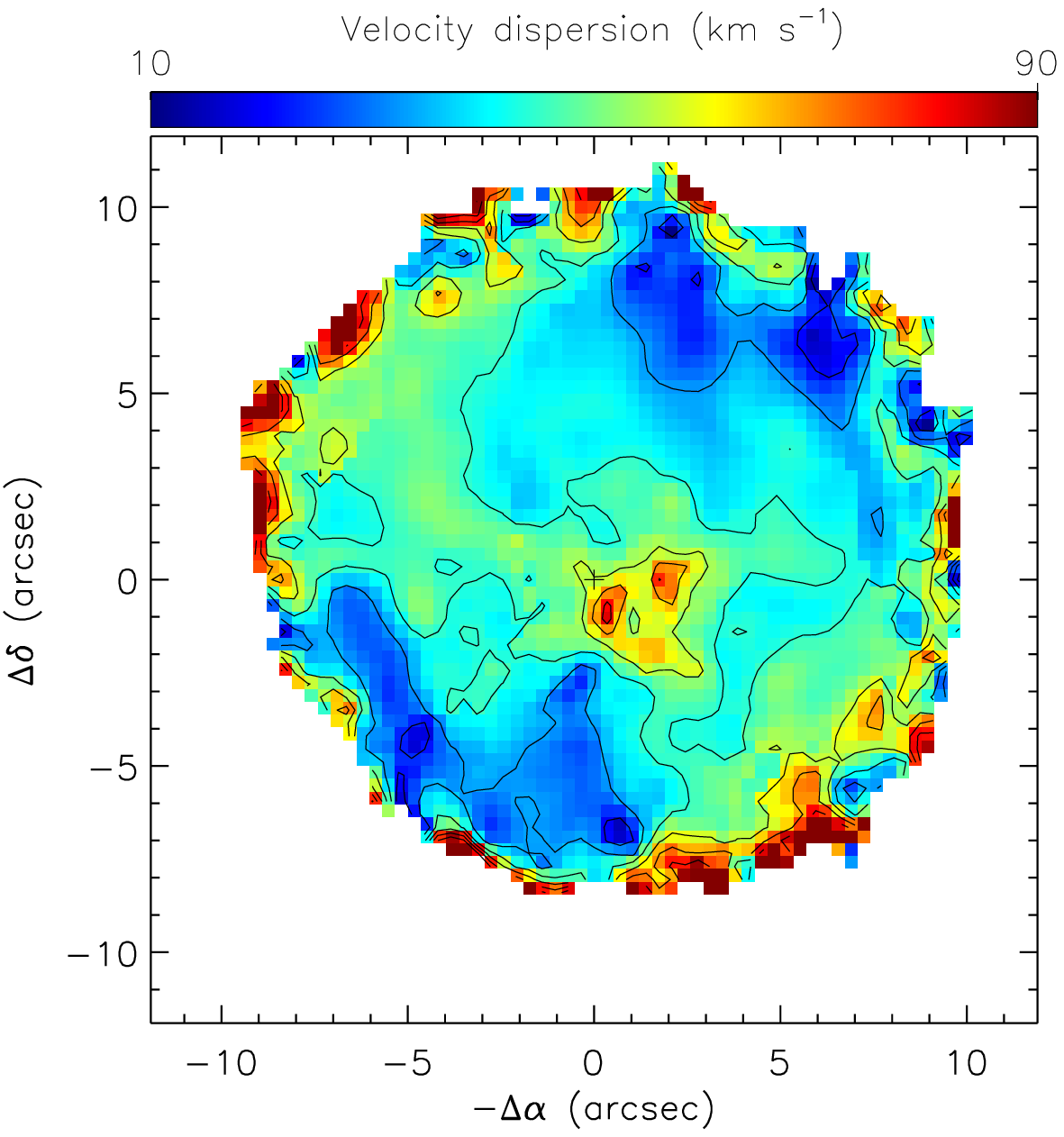}
}
\caption{FPI observations of Arp~148 (top) and II~Zw~28 (bottom).
Left panels: the H$\alpha$ images.
Middle panels: the velocity fields of the ionized gas.
Right panels: the maps of the ionized gas velocity dispersion.
}
\label{fig3}
\end{figure*}

We currently run a complex program of the FPI, long-slit and photometric
observations with the SAO RAS 6m telescope for several more prominent
northern collisional rings \citep{bizyaev09} in order to constrain the
evolution scenarii for the
ring galaxies from the comparison of their present-day and past kinematics.
We obtain patterns of the non-circular motions in the galaxies by subtracting
the model circular motion of ionized gas from the observed H$\alpha$ velocity
field. Our preliminary analysis of the collisional rings shown in Fig.~\ref{fig3}
suggests that the non-circular motions, including the radial
expansion, can be identified in all considered stellar systems. At the
present time, despite the valuable theoretical contributions to the
understanding of the nature of the collisional ring galaxies, the
observational confirmation of systematical radial motions is found
for just a few objects (Cartwheel and Arp~10, see references above) with
the help of radio HI or H$\alpha$ observations.
Kinematical observations of more number of the collisional rings will allow
us to investigate diversity of the internal motions in this kind of galaxies.

\section{Polar rings}

\subsection{Large-scale polar rings}

The polar-ring galaxies (PRGs) are objects which are determined by
the possession of the outer
rings or disks
of gas, dust and/or stars in them which orbit nearly orthogonally
to the plane of the main galaxy.
PRG are believed in most cases to be formed due to the galactic
mergers and accretion of the matter from a companion galaxy or gaseous
filaments from the intergalactic medium onto the host galaxy
(\citet{Bournaud03,Combes06} and references therein).

The circular rotation in two mutually perpendicular planes provides a rare
chance to probe the three-dimensional mass distribution in the galaxy
and to determine
the spatial shape of the dark halos \citep{Combes06}.
To do this, it is necessary to
obtain sufficiently detailed data about the inner and outer kinematics of
the PRG.
\citet{Whitmore90} list 157 candidates to the polar-ring galaxies
selected mostly by
their peculiar appearance. However, the number of true PRG, i.e. of those
which bona fide
exhibit a rotation in orthogonal planes, is much smaller, about twenty
galaxies. Even
in the simplest case when both the central galaxy and its ring are observed
edge-on, at least two long-slit spectroscopic sections are required to
determine the rotation pattern in both subsystems. A number of
problems arises in the case of objects where the host galaxy or ring is
observed with a
moderate inclination to the line-of-sight.
In these cases we have to determine the
two-dimensional velocity fields to study kinematic in such systems.
The example of NGC\,2655 is a good
illustration. It was suspected to have a polar ring due to the powerful dust
lane crossing the disk of the main galaxy.
The comparison of the velocity fields of
the gas and stars in the circumnuclear region confirmed this interpretation
\citep{SilAfanas2004}. Recently conducted 21-cm line observations allow
to analyze the structure
and kinematics of the polar ring  \citep{Sparke2008}.

One of the first FPI data for the PRGs were briefly presented by
\citet{Nicholson1987} who suggested a strong ($\sim90^\circ$) warp of the
outer regions of the polar gas disk in NGC~2685.  The detailed picture of
a complex helical warp was presented recently with HI data \citep{Jozsa2009}.
\citet{Sackett1995} made a conclusion about a significantly flattened
dark halo in A0136-0801 which is based on a relationship between the elongated
orbit position and FPI velocity field. Recently the St.Peterburg
University's team made an attempt to observe several polar rings with
panoramic spectroscopy devices and
the SAO RAS 6-m telescope \citep{Shalyapina2004,Hagen2005,Merkulova2008}.

\begin{figure*}
\centerline{
\includegraphics[width=5.5 cm]{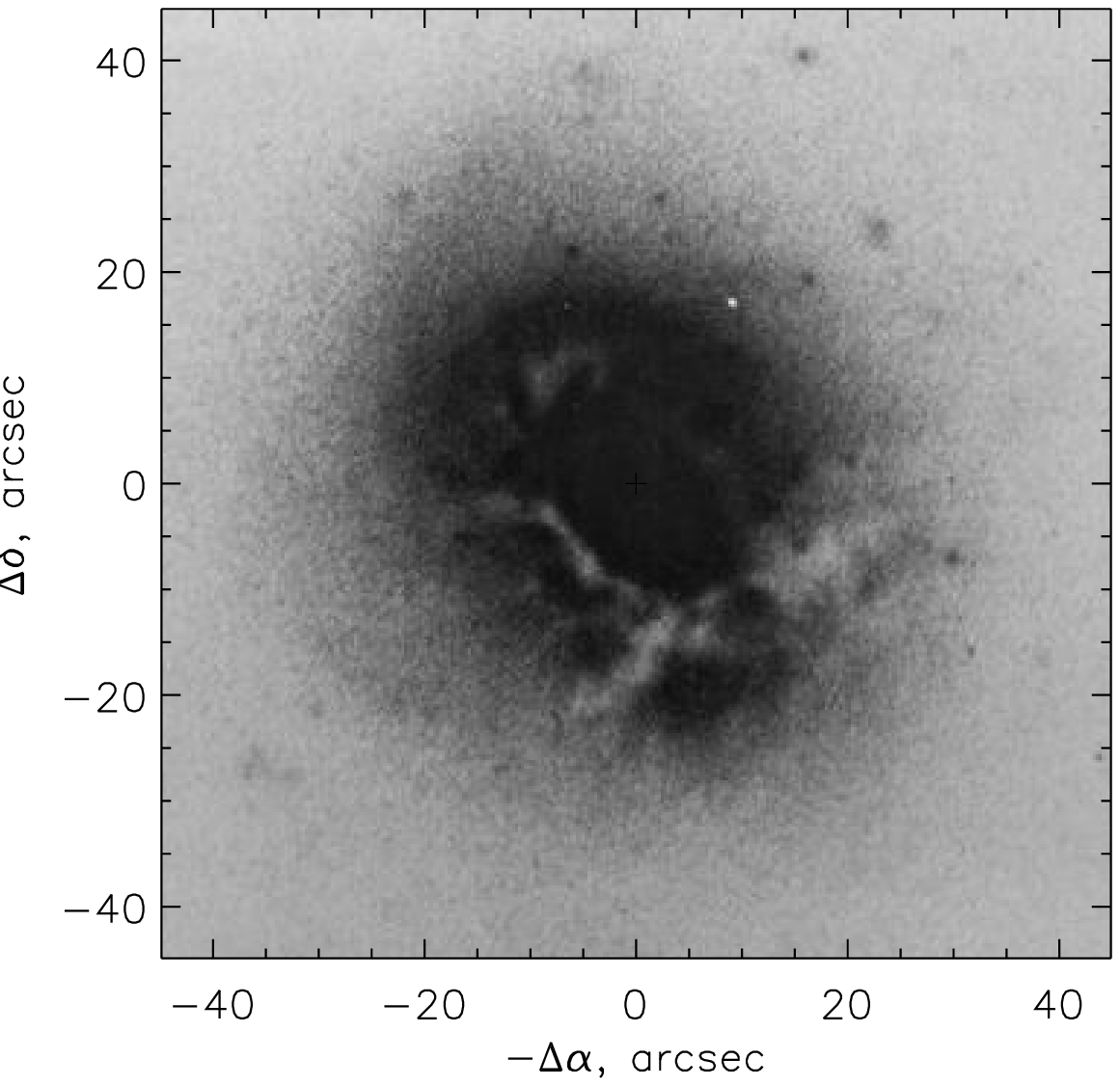}
\includegraphics[width=5.5 cm]{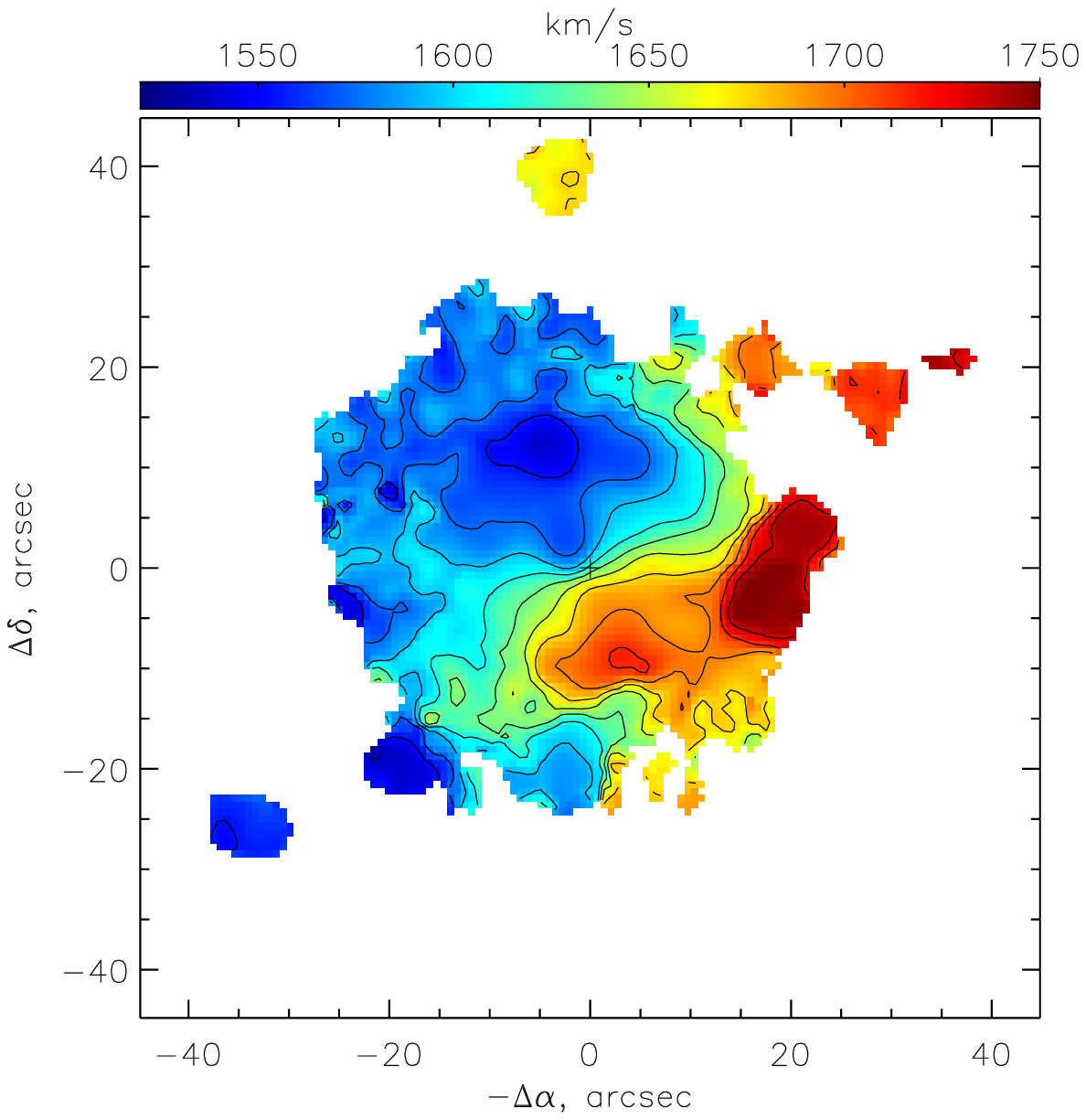}
\includegraphics[width=5.5 cm]{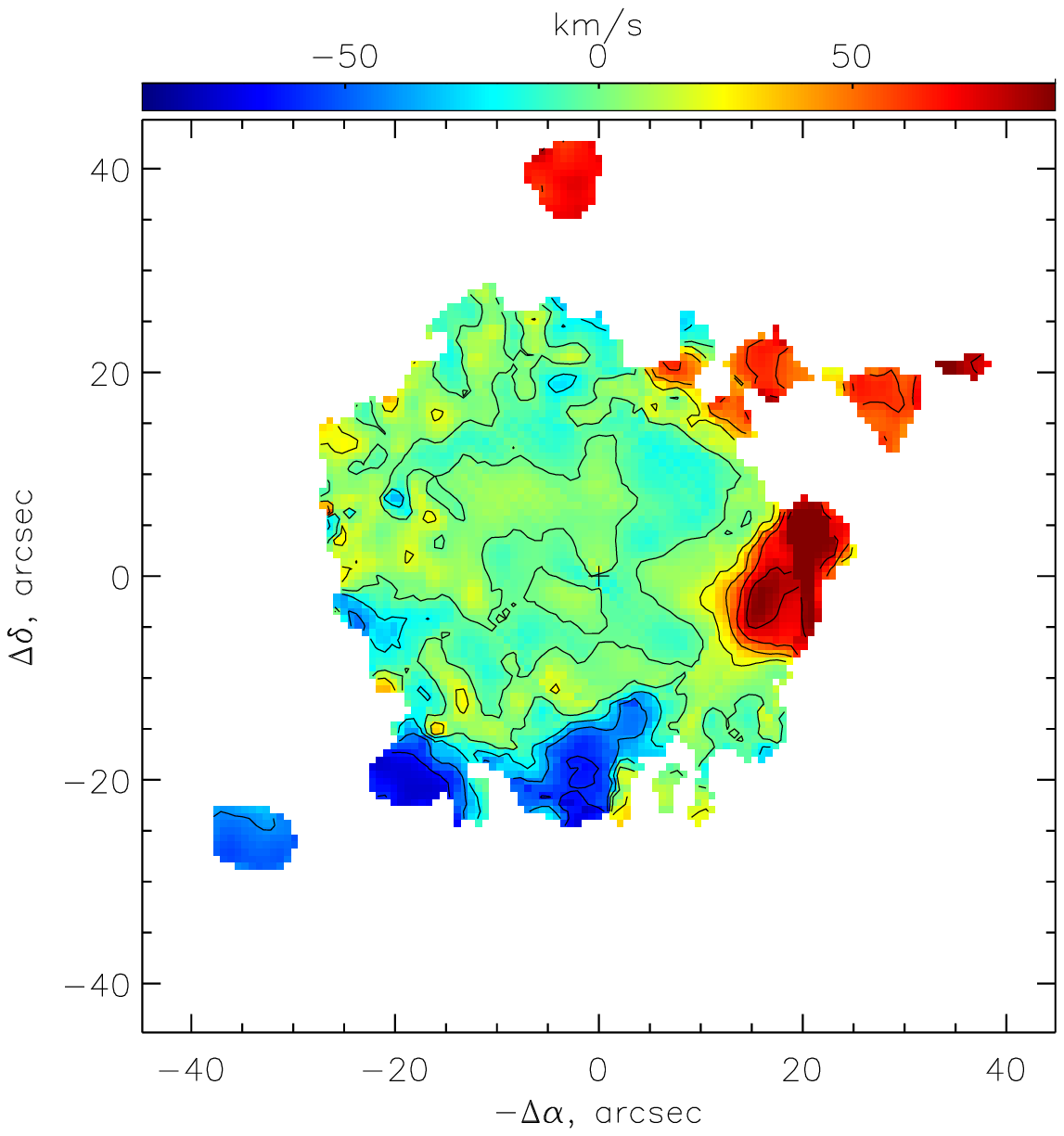}
}
 \caption{
Arp~212. Left: a red-plate photo  from the Arp's atlas. Middle: H$\alpha$
 velocity field obtained with the SCORPIO/FPI.
Right: the residual velocity field (after
subtraction of the model circular field found out from the rotation
of the inner disk)
reveals the kinematically decoupled regions which are seen as
knots in the warped polar ring}
\label{fig4}
\end{figure*}

Using the velocity fields enables to find a kinematically decoupled  polar
component even though it is not clearly distinguished in the morphological
structure. Recent observations of Arp~212 give a good example for it
\citep{Moiseev2008}. In the ionized gas, the two kinematically distinctive
subsystems are found (Fig.~\ref{fig4}). The first one, at $r<3.5$~kpc,
rotates in the plane of the stellar disk.  The second system is located at
$r=2-6$~ kpc from the center and consists of several isolated HII regions
whose orbits are tilted by high
angles to the main stellar disk. Our own observing
results and the data from literature on the
kinematics of molecular, ionized, and neutral gas can be explained in terms
of the model which implies that most of the gas in the outer parts of the galaxy
concentrates in a broad ring with radius of about 20~kpc. The outer ring
regions
rotate in the polar plane. Once the galactocentric distance decreases,
the orbits of the gaseous clouds precess and approach the galactic disk.
This precession is due to the non-spherical (maybe a triaxial) shape of the
gravitational potential in Arp~212.

\subsection{Nuclear polar rings and disks}

The 3D spectroscopy provides a good spatial sampling that is important for
studies
of the inner polar rings/disk with sizes from 100--200~pc to 1--2~kpc.
Their contribution to the total luminosity of the whole circumnuclear region
in their host galaxy
is small. Usually the ionized gas is mostly presented at the polar orbits,
however the mixed stellar-gaseous structures can also be observed.
\textit{It is interesting to
note that the number of confirmed inner polar structures is larger than
that of the `classical' PRGs.}
At the present time we count 31 objects with the
inner polar/inclined disks or rings. The sample includes 17 objects listed
by \citet{Corsini2003}, 8 galaxies studied by \citet{SilAfanas2004} and
a few more from other papers, see some references in \citet{Sil2009}.
The major fraction of the nuclear polar structures has been confirmed with
the 3D spectroscopy (integral-field or FPI) with the SAO RAS 6-m
telescope. Detailed observations reveal that in some cases the inner polar
orbits extend down to the equatorial plane at the large radii, as it
can be seen from the
VLT/VIMOS data for NGC~2685 \citep{Coccato2007}. The smooth transition from
the polar to the `normal' orientation of rotation is exhibited by the
ionized gas in late-type galaxy NGC~7468  observed with the SCORPIO/FPI
by \citet{Shalyapina2004}.

The origin of the inner polar structures is not well understood yet. The most
popular hypothesis   is the external gas accretion as for a
large-scale PRG.
However, some dynamical mechanisms of their formation are also proposed,
like a transportation of the
gas onto highly-inclined anomalous orbits by a tumbling triaxial potential,
see \citet{SilAfanas2004} for detailed discussion. More precise
observations and
numerical simulations are required to explain successfully these galactic
structures. We hope that increasing amounts of observational information
about the inner polar disks will stimulate new theoretical efforts
in this direction.

\medskip

This review is based on observations made with the 6-m telescope at the
Special Astrophysical Observatory of the Russian Academy of Sciences, which
is operated under the financial support of the Ministry of Science of the
Russian Federation (registration number 01-43). A.M. thanks the 7th SCSLSA
Organized Committee for their kind invitation and hosting. This work was
supported by the Russian Foundation for Basic Research (project
no.~09-02-00870)


\begin{thebibliography}{00}

\bibitem[{{Afanasiev} \& {Moiseev}(2005)}]{Afanasiev2005}
{Afanasiev}, V.~L., {Moiseev}, A.~V., 2005. Astronomy Letters 31, 194.

\bibitem[{{Amram} et~al.(1998){Amram}, {Mendes de Oliveira}, {Boulesteix}, \&
  {Balkowski}}]{amram98}
{Amram}, P., {Mendes de Oliveira}, C., {Boulesteix}, J., {Balkowski}, C., 1998.
  A\&A 330, 881.

\bibitem[{{Appleton} \& {Struck-Marcell}(1996)}]{appleton96}
{Appleton}, P.~N., {Struck-Marcell}, C., 1996. Fundamentals of Cosmic Physics
  16, 111.

\bibitem[{{Arsenault} et~al.(1988){Arsenault}, {Boulesteix}, {Georgelin}, \&
  {Roy}}]{Arsenault1988}
{Arsenault}, R., {Boulesteix}, J., {Georgelin}, Y., {Roy}, J.-R., 1988. A\&A
  200, 29.

\bibitem[{{Athanassoula} et~al.(1997){Athanassoula}, {Puerari}, \&
  {Bosma}}]{athanassoula97}
{Athanassoula}, E., {Puerari}, I., {Bosma}, A., 1997. MNRAS 286, 284.

\bibitem[{{Beckman} et~al.(2008){Beckman}, {Fathi}, {Pi{\~n}ol}, {Hernandez},
  {Carignan}, \& {P{\'e}rez}}]{Beckman2008}
{Beckman}, J., {Fathi}, K., {Pi{\~n}ol}, N., {Hernandez}, O., {Carignan}, C.,
  {P{\'e}rez}, I., 2008. To be published in MSAIt. arXiv:0810.3996 [astro-ph].

\bibitem[{{Bizyaev} et~al.(2009){Bizyaev}, {Moiseev}, \&   {Vorobyov}}]{bizyaev09}
{Bizyaev}, D., {Moiseev}, A., {Vorobyov}, E., 2009. In: Bulletin of the
  American Astronomical Society. Vol.~41. p. 328.

\bibitem[{{Bizyaev} et~al.(2007){Bizyaev}, {Moiseev}, \&   {Vorobyov}}]{bizyaev07}
{Bizyaev}, D.~V., {Moiseev}, A.~V., {Vorobyov}, E.~I., 2007. ApJ 662, 304.

\bibitem[{{Bournaud} \& {Combes}(2003)}]{Bournaud03}
{Bournaud}, F., {Combes}, F., 2003. A\&A 401, 817.

\bibitem[{{Buta}(1988)}]{Buta1988}
{Buta}, R., 1988. ApJs 66, 233.

\bibitem[{{Buta} et~al.(1998){Buta}, {Alpert}, {Cobb}, {Crocker}, \&   {Purcell}}]{Buta1998}
{Buta}, R., {Alpert}, A.~J., {Cobb}, M.~L., {Crocker}, D.~A., {Purcell}, G.~B.,
  1998. AJ 116, 1142.

\bibitem[{{Buta} \& {Combes}(1996)}]{butacombes96}
{Buta}, R., {Combes}, F., 1996. {Galactic Rings}. Fundamentals of Cosmic
  Physics 17, 95.

\bibitem[{{Buta} \& {Purcell}(1998)}]{ButaPurcell1998}
{Buta}, R., {Purcell}, G.~B., 1998. AJ 115, 484.

\bibitem[{{Buta} et~al.(1995){Buta}, {Purcell}, \& {Crocker}}]{Buta1995}
{Buta}, R., {Purcell}, G.~B., {Crocker}, D.~A., 1995. AJ 110, 1588.

\bibitem[{{Charmandaris} \& {Appleton}(1996)}]{charmandaris96}
{Charmandaris}, V., {Appleton}, P.~N., 1996. ApJ 460, 686.

\bibitem[{{Coccato} et~al.(2007){Coccato}, {Corsini}, {Pizzella}, \&   {Bertola}}]{Coccato2007}
{Coccato}, L., {Corsini}, E.~M., {Pizzella}, A., {Bertola}, F., 2007. A\&A 465,
  777.

\bibitem[{{Combes}(2006)}]{Combes06}
{Combes}, F., 2006. In: {Mamon}, G.~A., {Combes}, F., {Deffayet}, C., {Fort},
  B. (Eds.), EAS Publications Series. Vol.~20. p.~97.

\bibitem[{{Comer{\'o}n} et~al.(2008){Comer{\'o}n}, {Knapen}, {Beckman}, \&   {Shlosman}}]{comeron08}
{Comer{\'o}n}, S., {Knapen}, J.~H., {Beckman}, J.~E., {Shlosman}, I., Feb.
  2008. A\&A 478, 403.

\bibitem[{{Corsini} et~al.(2003){Corsini}, {Pizzella}, {Coccato}, \&   {Bertola}}]{Corsini2003}
{Corsini}, E.~M., {Pizzella}, A., {Coccato}, L., {Bertola}, F., 2003. A\&A 408,
  873.

\bibitem[{{Fathi} et~al.(2007){Fathi}, {Beckman}, {Zurita}, {Rela{\~n}o},
  {Knapen}, {Daigle}, {Hernandez}, \& {Carignan}}]{Fathi2007}
{Fathi}, K., {Beckman}, J.~E., {Zurita}, A., {Rela{\~n}o}, M., {Knapen}, J.~H.,
  {Daigle}, O., {Hernandez}, O., {Carignan}, C., 2007. A\&A 466, 905.

\bibitem[{{Fosbury} \& {Hawarden}(1977)}]{fosbury77}
{Fosbury}, R.~A.~E., {Hawarden}, T.~G., 1977. MNRAS 178, 473.

\bibitem[{{Ghosh} \& {Mapelli}(2008)}]{ugc7069}
{Ghosh}, K.~K., {Mapelli}, M., May 2008. MNRAS 386, L38.

\bibitem[{{Hagen-Thorn} et~al.(2005){Hagen-Thorn}, {Shalyapina}, {Karataeva},
  {Yakovleva}, {Moiseev}, \& {Burenkov}}]{Hagen2005}
{Hagen-Thorn}, V.~A., {Shalyapina}, L.~V., {Karataeva}, G.~M., {Yakovleva},
  V.~A., {Moiseev}, A.~V., {Burenkov}, A.~N., 2005. Astronomy Reports 49, 958.

\bibitem[{{Hernandez} et~al.(2005){Hernandez}, {Wozniak}, {Carignan}, {Amram},
  {Chemin}, \& {Daigle}}]{Hernandez2005}
{Hernandez}, O., {Wozniak}, H., {Carignan}, C., {Amram}, P., {Chemin}, L.,
  {Daigle}, O., 2005. ApJ 632, 253.

\bibitem[{{Higdon}(1996)}]{higdon96}
{Higdon}, J.~L., 1996. ApJ 467, 241.

\bibitem[{{J{\'o}zsa} et~al.(2009){J{\'o}zsa}, {Oosterloo}, {Morganti},
  {Klein}, \& {Erben}}]{Jozsa2009}
{J{\'o}zsa}, G.~I.~G., {Oosterloo}, T.~A., {Morganti}, R., {Klein}, U.,
  {Erben}, T., 2009. A\&A 494, 489.

\bibitem[{{Knapen} et~al.(2004){Knapen}, {Whyte}, {de Blok}, \& {van der
  Hulst}}]{Knapen2004}
{Knapen}, J.~H., {Whyte}, L.~F., {de Blok}, W.~J.~G., {van der Hulst}, J.~M.,
  2004. A\&A 423, 481.

\bibitem[{{Korchagin} et~al.(1998){Korchagin}, {Mayya}, {Vorobyov}, \&   {Kembhavi}}]{korchagin98}
{Korchagin}, V., {Mayya}, Y.~D., {Vorobyov}, E.~I., {Kembhavi}, A.~K., 1998.
  ApJ 495, 757.

\bibitem[{{Lin} et~al.(2008){Lin}, {Yuan}, \& {Buta}}]{Lin2008}
{Lin}, L.-H., {Yuan}, C., {Buta}, R., 2008. ApJ 684, 1048.

\bibitem[{{Lynds} \& {Toomre}(1976)}]{lynds76}
{Lynds}, R., {Toomre}, A., 1976. ApJ 209, 382.

\bibitem[{{Madore} et~al.(2009){Madore}, {Nelson}, \& {Petrillo}}]{Madore09}
{Madore}, B.~F., {Nelson}, E., {Petrillo}, K., 2009. ApJs 181, 572.

\bibitem[{{Mayya} et~al.(2005){Mayya}, {Bizyaev}, {Romano}, {Garcia-Barreto},
  \& {Vorobyov}}]{mayya05}
{Mayya}, Y.~D., {Bizyaev}, D., {Romano}, R., {Garcia-Barreto}, J.~A.,
  {Vorobyov}, E.~I., 2005. ApJl 620, L35.

\bibitem[{{Mazzei} et~al.(1995){Mazzei}, {Curir}, \& {Bonoli}}]{mazzei95}
{Mazzei}, P., {Curir}, A., {Bonoli}, C., 1995. AJ 110, 559.

\bibitem[{{Mazzuca} et~al.(2006){Mazzuca}, {Sarzi}, {Knapen}, {Veilleux}, \&   {Swaters}}]{Mazzuca2006}
{Mazzuca}, L.~M., {Sarzi}, M., {Knapen}, J.~H., {Veilleux}, S., {Swaters}, R.,
  2006. ApJl 649, L79.

\bibitem[{{Merkulova} et~al.(2008){Merkulova}, {Shalyapina}, {Yakovleva}, \&   {Karataeva}}]{Merkulova2008}
{Merkulova}, O.~A., {Shalyapina}, L.~V., {Yakovleva}, V.~A., {Karataeva},
  G.~M., 2008. Astronomy Letters 34, 542.

\bibitem[{{Moiseev}(2008)}]{Moiseev2008}
{Moiseev}, A.~V., 2008. Astrophysical Bulletin 63, 201. arXiv: 0808.1696 [astro-ph]

\bibitem[{{Moiseev} et~al.(2004){Moiseev}, {Vald{\'e}s}, \&   {Chavushyan}}]{Moiseev2004}
{Moiseev}, A.~V., {Vald{\'e}s}, J.~R., {Chavushyan}, V.~H., 2004. A\&A 421,
  433.

\bibitem[{{Mu{\~n}oz-Tu{\~n}{\'o}n} et~al.(2004){Mu{\~n}oz-Tu{\~n}{\'o}n},
  {Caon}, \& {Aguerri}}]{Casiana2004}
{Mu{\~n}oz-Tu{\~n}{\'o}n}, C., {Caon}, N., {Aguerri}, J.~A.~L., 2004. AJ 127,
  58.

\bibitem[{{Nicholson} et~al.(1987){Nicholson}, {Taylor}, {Sparks}, \&   {Bland}}]{Nicholson1987}
{Nicholson}, R.~A., {Taylor}, K., {Sparks}, W.~B., {Bland}, J., 1987. In: {de
  Zeeuw}, P.~T. (Ed.), Structure \& Dynamics of Elliptical Galaxies. Vol. 127
  of IAU Symposium. p. 415.

\bibitem[{{Sackett} \& {Pogge}(1995)}]{Sackett1995}
{Sackett}, P.~D., {Pogge}, R.~W., 1995. In: {Holt}, S.~S., {Bennett}, C.~L.
  (Eds.), Dark Matter. Vol. 336 of American Institute of Physics Conference
  Series. p. 141.

\bibitem[{{Salo} et~al.(1999){Salo}, {Rautiainen}, {Buta}, {Purcell}, {Cobb},
  {Crocker}, \& {Laurikainen}}]{Salo1999}
{Salo}, H., {Rautiainen}, P., {Buta}, R., {Purcell}, G.~B., {Cobb}, M.~L.,
  {Crocker}, D.~A., {Laurikainen}, E., 1999. AJ 117, 792.

\bibitem[{{Shalyapina} et~al.(2004){Shalyapina}, {Moiseev}, {Yakovleva},
  {Hagen-Thorn}, \& {Barsunova}}]{Shalyapina2004}
{Shalyapina}, L.~V., {Moiseev}, A.~V., {Yakovleva}, V.~A., {Hagen-Thorn},
  V.~A., {Barsunova}, O.~Y., 2004. Astronomy Letters 30, 583.

\bibitem[{{Sil'chenko} \& {Afanasiev}(2004)}]{SilAfanas2004}
{Sil'chenko}, O.~K., {Afanasiev}, V.~L., 2004. AJ 127, 2641.

\bibitem[{{Sil'chenko} \& {Moiseev}(2006)}]{silmois06}
{Sil'chenko}, O.~K., {Moiseev}, A.~V., 2006. AJ 131, 1336.

\bibitem[{{Sil'chenko} et~al.(2009){Sil'chenko}, {Moiseev}, \&   {Afanasiev}}]{Sil2009}
{Sil'chenko}, O.~K., {Moiseev}, A.~V., {Afanasiev}, V.~L., 2009. ApJ 694,
  1550.

\bibitem[{Smirnova et~al.(2006){Smirnova}, {Moiseev}, \&   {Afanasiev}}]{Smirnova2006}
{Smirnova}, A.~A., {Moiseev}, A.~V., {Afanasiev}, V.~L., 2006. Astronomy
  Letters 32, 520.

\bibitem[{{Sparke} et~al.(2008){Sparke}, {van Moorsel}, {Erwin}, \&   {Wehner}}]{Sparke2008}
{Sparke}, L.~S., {van Moorsel}, G., {Erwin}, P., {Wehner}, E.~M.~H., 2008. AJ
  135, 99.

\bibitem[{{Struck-Marcell} \& {Higdon}(1993)}]{struck93}
{Struck-Marcell}, C., {Higdon}, J.~L., 1993. ApJ 411, 108.

\bibitem[{{Theys} \& {Spiegel}(1976)}]{theys76}
{Theys}, J.~C., {Spiegel}, E.~A., 1976. ApJ 208, 650.

\bibitem[{{Theys} \& {Spiegel}(1977)}]{theys77}
{Theys}, J.~C., {Spiegel}, E.~A., 1977. ApJ 212, 616.

\bibitem[{{Tremaine} \& {Weinberg}(1984)}]{TW1984}
{Tremaine}, S., {Weinberg}, M.~D., 1984. ApJl 282, L5.

\bibitem[{Tutukov \& Fedorova(2006)}]{Tutukov2006}
{Tutukov}, A.~V., {Fedorova}, A.~V., 2006. Astronomy Reports 50, 785.

\bibitem[{{van der Kruit}(1974)}]{vanderKruit1974}
{van der Kruit}, P.~C., 1974. ApJ 188, 3.

\bibitem[{Vorobyov \& Bizyaev(2003)}]{vorobyov03}
{Vorobyov}, E.~I., {Bizyaev}, D., 2003. A\&A 400, 81.

\bibitem[{{Whitmore} et~al.(1990){Whitmore}, {Lucas}, {McElroy},
  {Steiman-Cameron}, {Sackett}, \& {Olling}}]{Whitmore90}
{Whitmore}, B.~C., {Lucas}, R.~A., {McElroy}, D.~B., {Steiman-Cameron}, T.~Y.,
  {Sackett}, P.~D., {Olling}, R.~P., 1990. AJ 100, 1489.



\end{thebibliography}
\end{document}